\documentstyle[12pt]{article}
\begin{document}

\vskip 2cm

\begin{center}

{\Large {\sf Dual-BRST symmetry: 6D Abelian 3-form gauge theory}}

\vskip 2.5 cm

{\sf{ R. Kumar$^{(a)}$, S. Krishna$^{(a)}$, A. Shukla$^{(a)}$, R. P. Malik$^{(a,b)}$}}\\
{\it $^{(a)}$Department of Physics, Centre of Advanced Studies, Faculty of Science,}\\
{\it Banaras Hindu University, Varanasi - 221 005, (Uttar Pradesh), India}\\

\vskip 0.1cm

{\bf and}\\

\vskip 0.1cm

{\it $^{(b)}$DST Centre for Interdisciplinary Mathematical Sciences,}\\
{\it Faculty of Science, Banaras Hindu University, Varanasi - 221 005, India}\\
{\small {\sf {E-mails: raviphynuc@gmail.com; skrishna.bhu@gmail.com; ashukla038@gmail.com; malik@bhu.ac.in}}}
\end{center}

\vskip 2cm

\noindent
{\bf Abstract:} 
Within the framework of Becchi-Rouet-Stora-Tyutin (BRST) formalism, we demonstrate the existence of the novel
off-shell nilpotent (anti-)dual-BRST symmetries in the context of a six (5 + 1)-dimensional (6D) free Abelian 
3-form gauge theory.  Under these local and continuous symmetry transformations, the total gauge-fixing term of the Lagrangian density 
remains invariant. This observation should be contrasted with the off-shell nilpotent (anti-)BRST symmetry transformations, 
under which, the total kinetic term of the theory remains invariant. The anticommutator of the above nilpotent (anti-)BRST and 
(anti-)dual-BRST transformations leads to the derivation of a bosonic symmetry in the theory. There exists a discrete symmetry transformation
in the theory which provides a thread of connection between the nilpotent (anti-)BRST and (anti-)dual-BRST  transformations. This theory is 
endowed with a ghost-scale symmetry, too.  We discuss the algebra of these symmetry transformations and show that the structure of the algebra
is reminiscent of the algebra of  de Rham cohomological operators of differential geometry. \\



\newpage

\noindent
\section {Introduction} 
The study of the higher dimensional ($D>4$) and higher $p$-form ($p\geq 2$)
gauge theories is important because
these higher $p$-form ($p = 2, 3,...$) fields appear in the excitations of the quantized versions of 
(super)strings and related extended objects
(see, e.g., [1-3] for details). Furthermore, it is now well-established that the consistent quantum theories of these extended objects
do live in dimensions much higher than the ordinary 4D spacetime. Thus, the modern developments in (super)string (and related extended objects) have
spurred the interest of theorists to study field theories that are connected with the higher $p$-form gauge theories in higher $(D > 4)$ dimensions 
of spacetime. In some sense, these field theories are generalizations of the usual field theories in the ordinary 4D spacetime.
With this background in mind, our present endeavor is an attempt to explore the interesting symmetries associated with the free 
Abelian 3-form gauge theory in six dimensions of spacetime.

In addition to the above motivation, there is another argument in favor of our interest in the study of higher
$p$-form ($p \geq 2$) gauge theories. For instance, it has been shown [4-6] that a 4D Abelian 1-form gauge field 
acquires a mass without any recourse to the Higgs mechanism when it is merged with an Abelian 2-form 
[$B^{(2)} = \frac {1}{2!} (dx^\mu \wedge dx^\nu)B_{\mu\nu}$] antisymmetric gauge field ($B_{\mu\nu}$)
through the celebrated topological ($B \wedge F$) term. Attempts to generalize this model
to the topologically massive 4D non-Abelian gauge theory have been made in the past [7-10]. Furthermore,
these models have also been studied within the framework of Becchi-Rouet-Stora-Tyutin (BRST) formalism in our earlier works [11-14]. In exactly above fashion, it has been shown that the merging of Abelian 2-form and 3-form gauge
fields, through a topological term in the 6D spacetime, leads to the mass generation of the Abelian 2-form 
gauge field (see, e.g., [15]).
Thus, it is important to explore the details of the higher $p$-form ($p\geq 2$) gauge theories because these provide an 
alternative to the Higgs mechanism as far as the mass generation of gauge fields is concerned.

There is yet another reason behind our interest in the present investigation.
In a very recent paper [16], we have claimed that there always exists a set of proper (anti-)dual-BRST symmetries, 
for an
Abelian $p$-form gauge theory in $D = 2p$ dimensions of spacetime, within the framework of BRST formalism.
We have already shown in our earlier works on 2D Abelian 1-form gauge theory [17-19] and 4D Abelian 2-form
gauge theory [20-22] that the above nilpotent and absolutely anticommuting (anti-)co-BRST symmetries do exist.
The purpose of our present investigation is to corroborate the above claim by demonstrating that the proper
(i.e. nilpotent and absolutely anticommuting) 
(anti-)dual-BRST symmetries do exist for the 6D Abelian 3-form gauge theory
 under which the total gauge-fixing term, owing its origin to the co-exterior derivative 
of differential geometry [23-25], remains invariant.

In our present investigation, we demonstrate that there exist six continuous symmetries
and one discrete symmetry in the theory which obey the algebra of the de Rham cohomological operators
of differential geometry. In particular, an elegant interplay between the discrete and continuous
symmetry transformations provides a physical realization of the relationship between the exterior and co-exterior
derivatives of differential geometry.
As a consequence, our present model is a field theoretic realization of the Hodge theory because all
the de Rham cohomological operators and their relationships are realized in the language of 
symmetry transformations on the relevant fields of our present theory.

Our present paper is organized as follows. In Sec. 2, we recapitulate the bare essentials of our earlier work [26] to 
set up the notations and conventions corresponding to the (anti-)BRST symmetry transformations. Our Sec. 3 is devoted
to the discussion of the (anti-)dual-BRST symmetry transformations. In Sec. 4, we deal with the derivation of the bosonic symmetry
from the anticommutator of the (anti-)BRST and (anti-)co-BRST symmetry transformations. We discuss the existence of ghost-scale
and discrete symmetries of the theory in Sec. 5. In the next Sec. 6, we deduce
the algebraic structures of all the above symmetry transformations. Finally, we make some concluding remarks in Sec. 7.

\noindent
\section {Preliminaries: (anti-)BRST symmetries}

Let us begin with the following coupled (but equivalent) Lagrangian densities for the free 6D Abelian 
3-form gauge theory\footnote{We adopt here the convention and notations such that the 6D spacetime
background Minkowskian manifold is endowed with a flat metric with signatures $(+1, -1, -1, -1, -1, -1)$
and corresponding totally antisymmetric Levi-Civita  tensor $\varepsilon_{\mu\nu\eta\kappa\rho\sigma}$
obeys the relations: $\varepsilon_{\mu\nu\eta\kappa\rho\sigma}\varepsilon^{\mu\nu\eta\kappa\rho\sigma} = - 6!,
\varepsilon_{\mu\nu\eta\kappa\rho\sigma}\varepsilon^{\mu\nu\eta\kappa\rho\lambda} = - 5! \delta^\lambda_\sigma,$ etc., 
with $\varepsilon_{012345} = + 1$ and the Greek indices $\mu, \nu, \eta,... = 0, 1, 2, 3, 4, 5$ correspond to the 6D spacetime directions. 
For the sake of brevity, we also use the notation $A\cdot B = A_\mu B^\mu = A_0 B_0 - A_i B_i$ where the Latin indices 
$i, j, k,... = 1, 2, 3, 4, 5$ correspond to only the space directions .} [26]
\begin{eqnarray}
{\cal L}_B &=& \frac{1}{24} H^{\mu\nu\eta\xi} H_{\mu\nu\eta\xi}
+  B^{\mu\nu} \Bigl ( \partial^\eta A_{\eta\mu\nu} + \frac{1}{2} 
[\partial_\mu \phi^{(1)}_\nu - \partial_\nu \phi^{(1)}_\mu] \Bigr )
 - \frac{1}{2} B_{\mu\nu}  B^{\mu\nu}\nonumber\\
&+& (\partial_\mu \bar C_{\nu\eta} + \partial_\nu \bar C_{\eta\mu} 
+ \partial_\eta \bar C_{\mu\nu})(\partial^\mu C^{\nu\eta}) + (\partial \cdot \phi^{(1)}) B_1 - \frac{1}{2} B_1^2 - B B_2\nonumber\\
&-& (\partial_\mu \bar \beta_\nu - \partial_\nu \bar \beta_\mu) (\partial^\mu \beta^\nu) 
+ (\partial_\mu \bar C^{\mu\nu} + \partial^\nu \bar C_1) f_\nu  -  (\partial_\mu  C^{\mu\nu} 
- \partial^\nu C_1) \bar F_\nu \nonumber\\
&+& \partial_\mu \bar C_2 \partial^\mu C_2 + (\partial \cdot \beta) B_2 - (\partial \cdot \bar \beta) B,
\end{eqnarray}
\begin{eqnarray}
{\cal L}_{\bar B} &=& \frac{1}{24} H^{\mu\nu\eta\xi} H_{\mu\nu\eta\xi}
-  {\bar B}^{\mu\nu} \Bigl ( \partial^\eta A_{\eta\mu\nu} - \frac{1}{2} 
[\partial_\mu \phi^{(1)}_\nu - \partial_\nu \phi^{(1)}_\mu] \Bigr )
 - \frac{1}{2} {\bar B}_{\mu\nu} {\bar B}^{\mu\nu}\nonumber\\
 &+& (\partial_\mu \bar C_{\nu\eta} + \partial_\nu \bar C_{\eta\mu} 
+ \partial_\eta \bar C_{\mu\nu}) (\partial^\mu C^{\nu\eta}) + (\partial \cdot \phi^{(1)}) B_1 - \frac{1}{2} B_1^2 - B B_2 \nonumber\\
&-& (\partial_\mu \bar \beta_\nu - \partial_\nu \bar \beta_\mu) (\partial^\mu \beta^\nu) 
- (\partial_\mu \bar C^{\mu\nu} + \partial^\nu \bar C_1) F_\nu + (\partial_\mu  C^{\mu\nu} 
- \partial^\nu C_1) \bar f_\nu \nonumber\\
&+& \partial_\mu \bar C_2 \partial^\mu C_2 + (\partial \cdot \beta) B_2 - (\partial \cdot \bar \beta) B,
\end{eqnarray}
where  $H_{\mu\nu\eta\kappa} = \partial_\mu A_{\nu\eta\kappa} -\partial_\nu A_{\eta\kappa\mu}+
\partial_\eta A_{\kappa\mu\nu}-\partial_\kappa A_{\mu\nu\eta}$
is the totally antisymmetric curvature tensor derived from the 4-form 
$H^{(4)} = (1/4!)(dx^\mu \wedge dx^\nu \wedge dx^\eta \wedge dx^\kappa)
H_{\mu\nu\eta\kappa} = dA^{(3)}$. The latter is obtained from the operation of the ordinary exterior derivative 
$d = dx^\mu \partial_\mu $ (with $d^2 = 0$) on the connection 3-form  [$A^{(3)} = (1/3!)(dx^\mu \wedge dx^\nu \wedge dx^\eta) A_{\mu\nu\eta}$] 
which defines the totally antisymmetric tensor Abelian 3-form gauge field $ A_{\mu\nu\eta}$.
The gauge-fixing term $(\partial^\eta A_{\eta\mu\nu})$ owes its origin to the co-exterior derivative 
$\delta = - * d*$ because the 2-form $\delta A^{(3)} = (1/2!) (dx^\mu \wedge dx^\nu)(\partial^\eta A_{\eta\mu\nu})$ captures it.
In the above discussion, the exterior and co-exterior derivatives are related to each-other by the Hodge duality $*$ operation. It should 
be noted that one has the freedom to add/subtract a 2-form ($F^{(2)}$) in the gauge-fixing term ($\partial^\eta A_{\eta\mu\nu}$). 
This can be easily done by applying an exterior derivative $d = dx^\mu \partial_\mu$ on a 1-form 
$\Phi^{(1)} = dx^\mu \phi^{(1)}_\mu$. This is precisely the reason that we have incorporated the vector field $\phi^{(1)}_\mu$ from
$ F^{(2)} = (1/2!)(dx^\mu \wedge dx^\nu) (\partial_\mu \phi^{(1)} _\nu - \partial_\nu \phi^{(1)}_\mu) = d \Phi^{(1)}$ in the coupled Lagrangian densities 
${\cal L}_{B}$ and ${\cal L}_{\bar B}$ that are associated with the linearization of the gauge-fixing terms of the theory.

In the above equations (1) and (2),  $({\bar C}_{\mu\nu})C_{\mu\nu}$ are the fermionic antisymmetric 
($\bar C_{\mu\nu} = - \bar C_{\nu\mu}, C_{\mu\nu} = - C_{\nu\mu}$) (anti-)ghost 
fields with ghost numbers $(-1)+1$, $({\bar \beta}_\mu)\beta_\mu$ are the bosonic ghost-for-ghost 
(anti-)ghost Lorentz vector fields with ghost numbers $(-2)+2$, (${\bar C}_2$)$C_2$ are the fermionic 
ghost-for-ghost-for-ghost (anti-)ghost Lorentz scalar fields with ghost numbers $(-3)+3$, respectively. 
Furthermore, we have antisymmetric Nakanishi-Lautrup type bosonic auxiliary fields ($B_{\mu\nu},{\bar B}_{\mu\nu}$) 
as well as the Lagrange multiplier fields $B, B_1$ and $B_2$ in the theory which have been invoked for the purpose of linearization of  
some specific terms. Similarly, we have $({\bar f}_\mu)f_\mu$ and $({\bar F}_\mu)F_\mu$ in our present theory as the fermionic auxiliary 
(anti-)ghost fields with ghost numbers ($-1$)$+1$, respectively. In the complete theory, 
we have fermionic (anti-)ghost fields (${\bar C}_1$)$C_1$ [with ghost numbers $(-1)+1$] and bosonic auxiliary  ghost fields $B_2$ and $B$ 
(which carry the ghost numbers ($-2$)$+2$), respectively.

It can be checked that, under the following off-shell nilpotent ($s_b^2 = 0$) BRST symmetry transformations ($s_b$)
(see, e.g. [26] for details):
\begin{eqnarray}
&&s_b A_{\mu\nu\eta} = \partial_\mu C_{\nu\eta} + \partial_\nu C_{\eta\mu}
+ \partial_\eta C_{\mu\nu}, \qquad s_b C_{\mu\nu} = \partial_\mu \beta_\nu
- \partial_\nu \beta_\mu, \qquad s_b \bar C_{\mu\nu} = B_{\mu\nu}, \nonumber\\
&&s_b \bar B_{\mu\nu} = \partial_\mu f_\nu - \partial_\nu f_\mu, \;\qquad
s_b \bar \beta_\mu = \bar F_\mu, \;\qquad
s_b \beta_\mu = \partial_\mu C_2, \;\qquad s_b F_\mu = - \partial_\mu B, \nonumber\\
&&s_b {\bar C}_2 = B_2, \qquad s_b C_1 = - B, \qquad s_b \bar C_1 = B_1, \qquad s_b \phi^{(1)}_\mu = f_\mu, 
\qquad s_b \bar f_\mu = \partial_\mu B_1,\nonumber\\
&&s_b \; \Bigl [ C_2,\; f_\mu,\; {\bar F}_\mu,\; B,\; B_1,\; B_2,\; B_{\mu\nu}, \;H_{\mu\nu\eta\kappa} \;\Bigl ]
\; = \;0,
\end{eqnarray}
the Lagrangian density ${\cal L}_B$ transforms to a total spacetime derivative as given below
\begin{eqnarray}
s_b {\cal L}_ B &=&  \partial_\mu \Bigl [ (\partial^\mu C^{\nu\eta} + \partial^\nu C^{\eta\mu}
+ \partial^\eta C^{\mu\nu})  B_{\nu\eta} + B^{\mu\nu} f_\nu 
- (\partial^\mu \beta^\nu - \partial^\nu \beta^\mu) \bar F_\nu \nonumber\\
&+& B_1 f^\mu - B \bar F^\mu + B_2 \partial^\mu C_2 \Bigr ].
\end{eqnarray}
In an exactly similar fashion, under the following off-shell nilpotent  ($s_{ab}^2 = 0$) anti-BRST 
symmetry transformations ($s_{ab}$) (see, e.g. [26] for details)
\begin{eqnarray}
&&s_{ab} A_{\mu\nu\eta} = \partial_\mu \bar C_{\nu\eta} + \partial_\nu \bar C_{\eta\mu}
+ \partial_\eta \bar C_{\mu\nu}, \quad  s_{ab} \bar C_{\mu\nu} = \partial_\mu \bar \beta_\nu
- \partial_\nu \bar \beta_\mu, \quad s_{ab}  C_{\mu\nu} = \bar B_{\mu\nu}, \nonumber\\
&&s_{ab} B_{\mu\nu} = \partial_\mu \bar f_\nu - \partial_\nu \bar f_\mu, \;\;\quad
s_{ab}  \beta_\mu =  F_\mu, \;\;\quad
s_{ab} \bar \beta_\mu = \partial_\mu \bar C_2, \;\quad s_{ab} \bar F_\mu = - \partial_\mu B_2, \nonumber\\
&&s_{ab} C_2 = B, \quad s_{ab} f_\mu = - \partial_\mu B_1, \quad s_{ab} C_1 = - B_1, \quad s_{ab} \bar C_1 = - B_2, 
\quad s_{ab} \phi^{(1)}_\mu = \bar f_\mu,\nonumber\\
&&s_{ab}\; \Bigl [ \bar C_2, \;\bar f_\mu, \;F_\mu,\; B,\; B_1, \;B_2, \;\bar B_{\mu\nu}, \;H_{\mu\nu\eta\kappa}
\;\Bigl ] \;= \;0,
\end{eqnarray}
the Lagrangian density ${\cal L}_{\bar B}$ transforms to a total spacetime derivative as illustrated below:
\begin{eqnarray}
s_{ab} {\cal L}_{\bar B} &=& \partial_\mu \Bigl [- (\partial^\mu {\bar C}^{\nu\eta} + \partial^\nu {\bar C}^{\eta\mu}
+ \partial^\eta {\bar C}^{\mu\nu}) \bar B_{\nu\eta} + {\bar B}^{\mu\nu} {\bar f}_\nu - (\partial^\mu {\bar \beta}^\nu -
\partial^\nu {\bar \beta}^\mu) F_\nu \nonumber\\
&+& B_1 {\bar f}^\mu + B_2 F^\mu - B \partial^\mu {\bar C}_2 \Bigr ].
\end{eqnarray}
As a consequence, the action integral ($S = \int d^6x {\cal L}_{(B, \bar B)}$) remains invariant for the
well-defined physical fields (incorporated in the theory) which vanish at infinity.

We close this section with the following comments. First, the above off-shell nilpotent BRST and anti-BRST 
symmetry transformations absolutely anticommute $(s_b s_{ab} + s_{ab} s_b = 0)$ with each-other only when the following CF-type restrictions [26,27]
\begin{eqnarray} 
f_\mu + F_\mu = \partial_\mu C_1, \qquad \bar f_\mu + \bar F_\mu = \partial_\mu \bar C_1, 
\qquad B_{\mu\nu} + \bar B_{\mu\nu} = \partial_\mu \phi^{(1)}_\nu - \partial_\nu \phi^{(1)}_\mu,
\end{eqnarray}
are satisfied. Second, the Lagrangian densities ${\cal L}_B$ and ${\cal L}_{\bar B}$ are equivalent (see, e.g. [26]) and both of 
them respect the (anti-)BRST symmetry transformations ($s_{(a)b}$) on a constraint hypersurface described 
by the CF-type restrictions given in (7). Third, the kinetic term [$(1/24) H^{\mu\nu\eta\kappa}H_{\mu\nu\eta\kappa}$], owing  its origin to the exterior derivative $d = dx^\mu \partial_\mu$, remains invariant under the (anti-)BRST symmetry transformations. Thus, one of the (anti-)BRST symmetry 
transformations provides a physical realization of $d$. Finally, the above (anti-)BRST symmetry transformations and CF-type restrictions have 
been derived from the superfield formalism discussed in [27] and they satisfy the key requirements of the nilpotency and
anticommutativity properties which are very sacrosanct  in the context of discussion of any  arbitrary $p$-form 
($p$ = 1, 2, 3,...) gauge theory
within the framework of BRST formalism.

\noindent
\section {(Anti-)dual-BRST symmetries}
As we have linearized the gauge-fixing term by introducing the Nakanishi-Lautrup field $ B_{\mu\nu}$ (and/or $\bar B_{\mu\nu}$)
and a vector field $\phi_\mu^{(1)}$ from the 
2-form $F^{(2)} = d\phi^{(1)} = (1/2!) (dx^\mu \wedge dx^\nu)(\partial_\mu \phi_\nu^{(1)} - \partial_\nu \phi_\mu^{(1)})$, 
similarly the kinetic term [$(1/24) H^{\mu\nu\eta\kappa}H_{\mu\nu\eta\kappa}$] can be linearized by invoking the 
auxiliary antisymmetric field ${\cal B}_{\mu\nu}$ (and/or ${\bar {\cal B}}_{\mu\nu}$) and $\phi_\mu^{(2)}$ from another 2-form 
${\tilde F}^{(2)} = d {\tilde\Phi}^{(1)} = (1/2!) (dx^\mu \wedge dx^\nu)(\partial_\mu \phi_\nu^{(2)} 
- \partial_\nu \phi_\mu^{(2)})$. Thus, the Lagrangian densities (1) and (2) can be linearized to the following 
\begin {eqnarray}
{\cal L}_{(B, {\cal B})} &=& \frac{1}{2} \;{\cal B}_{\mu\nu}\; {\cal B}^{\mu\nu}
 - {\cal B}^{\mu\nu} \Bigl(\frac{1}{4!} \; \varepsilon_{\mu\nu\eta\kappa\rho\sigma} H^{\eta\kappa\rho\sigma}
 + \frac{1}{2}\; [\partial_\mu \phi_\nu^{(2)} - \partial_\nu \phi_\mu^{(2)}]\Bigr) - \frac{1}{2}\;B^{\mu\nu} B_{\mu\nu}   \nonumber\\
&+& B^{\mu\nu} \Bigl(\partial^\eta A_{\eta\mu\nu} + \frac{1}{2} [\partial_\mu \phi_\nu^{(1)} 
- \partial_\nu \phi_\mu^{(1)}]\Bigr) + \Bigl(\partial_\mu \bar C_{\nu\eta} + \partial_\nu \bar C_{\eta\mu} + \partial_\eta \bar C_{\mu\nu} \Bigr)
 \Bigl (\partial^\mu C^{\nu\eta}\Bigr ) \nonumber\\ &+& (\partial \cdot \phi^{(1)})B_1 - (\partial \cdot \phi^{(2)})B_3
- \frac{1}{2}\; B_1^2 +\frac{1}{2}\;B^2_3 - (\partial \cdot\bar \beta) B - (\partial_\mu \bar \beta_\nu 
- \partial_\nu \bar \beta_\mu) (\partial^\mu \beta^\nu)\nonumber\\
& +& (\partial \cdot \beta) B_2 - B B_2 +(\partial_\mu \bar C^{\mu\nu} + \partial^\nu \bar C_1) f_\nu  -  (\partial_\mu  C^{\mu\nu} 
- \partial^\nu C_1) \bar F_\nu  + \partial_\mu \bar C_2 \partial^\mu C_2,
\end{eqnarray}
\begin {eqnarray}
{\cal L}_{(\bar B, {\bar{\cal B}})} &=& \frac{1}{2} \;\bar {\cal B}_{\mu\nu}\; \bar {\cal B}^{\mu\nu}
+ \bar {\cal B}^{\mu\nu} \Bigl(\frac{1}{4!} \; \varepsilon_{\mu\nu\eta\kappa\rho\sigma} H^{\eta\kappa\rho\sigma}
- \frac{1}{2}\; [\partial_\mu \phi_\nu^{(2)} - \partial_\nu \phi_\mu^{(2)} ]\Bigr) - \frac{1}{2}\;\bar B^{\mu\nu} \bar B_{\mu\nu}   \nonumber\\
&-& \bar B^{\mu\nu}\Bigl(\partial^\eta A_{\eta\mu\nu} - \frac{1}{2} [\partial_\mu \phi_\nu^{(1)} 
- \partial_\nu \phi_\mu^{(1)} ]\Bigr) + \Bigl(\partial_\mu \bar C_{\nu\eta} + \partial_\nu \bar C_{\eta\mu} + \partial_\eta \bar C_{\mu\nu}\Bigr)
\Bigl(\partial^\mu C^{\nu\eta}\Bigr ) \nonumber\\ &+& (\partial \cdot \phi^{(1)})B_1 - (\partial \cdot \phi^{(2)})B_3 -\frac{1}{2}\; B_1^2 +\frac{1}{2}\;B^2_3  -
(\partial \cdot\bar \beta) B - (\partial_\mu \bar \beta_\nu - \partial_\nu \bar \beta_\mu) (\partial^\mu \beta^\nu)\nonumber\\
&+& (\partial \cdot \beta) B_2 - B B_2 - (\partial_\mu \bar C^{\mu\nu} + \partial^\nu \bar C_1) F_\nu  +  (\partial_\mu  C^{\mu\nu} 
- \partial^\nu C_1) \bar f_\nu  + \partial_\mu \bar C_2 \partial^\mu C_2.
\end{eqnarray}
It should be noted that the novel gauge-fixing term [ i.e. $-\;(\partial \cdot \phi^{(2)})B_3 + \frac{1}2{} B_3^2$],
corresponding to the additional vector field $\phi_\mu^{(2)}$
has been incorporated through the Nakanishi-Lautrup type auxiliary field $B_3$
(in the above coupled Lagrangian densities) so that the theory, in its full blaze of glory, could became complete in all respects.

The following off-shell nilpotent ($s_d^2 = 0$) dual-BRST symmetry transformations ($s_d$):
\begin {eqnarray}
&&s_d A_{\mu\nu\eta} = \frac{1}{2}\; \varepsilon_{\mu\nu\eta\kappa\rho\sigma} \partial^{\kappa} {\bar C}^{\rho\sigma}, 
\quad s_d {\bar C}_{\mu\nu} = \partial_\mu {\bar \beta}_\nu - \partial_\nu {\bar \beta}_\mu, \quad 
s_d {\bar \beta}_\mu = \partial_\mu {\bar C}_2,  \nonumber\\ 
&&s_d {\bar C}_1 = - B_2, \quad s_d \beta_\mu = - f_\mu, \quad s_d C_1 = - B_3, \quad
s_d \phi_\mu^{(2)} = \bar F_\mu, \quad s_d C_2 = B,\nonumber\\
&&s_d C_{\mu\nu} = {\cal B}_{\mu\nu}, \quad s_d \bar f_\mu = \partial_\mu B_2,
\quad s_d {\bar {\cal B}}_{\mu\nu} = \partial_\mu \bar F_\nu -  \partial_\nu \bar F_\mu, \quad s_d F_\mu = \partial_\mu B_3 ,\nonumber\\
&& s_d \; [\partial^{\eta} A_{\eta\mu\nu}, \;\phi_\mu^{(1)},\; B_{\mu\nu}, \;{\cal B}_{\mu\nu},\;
B,\; B_1,\; B_2,\; B_3,\; \bar C_2,\; f_\mu,\; \bar F_\mu \;] \;= \;0,
\end{eqnarray}
leave the Lagrangian density ${\cal L}_{(B, {\cal B})}$ quasi-invariant. This is due to the fact that the latter transforms 
 to a total spacetime derivative, under $s_d$, as illustrated below:
\begin {eqnarray} 
s_d {\cal L}_{(B, {\cal B})} &=& - \partial_\mu \Bigl [(\partial^\mu \bar C^{\nu\eta} + \partial^\nu \bar C^{\eta\mu}
 + \partial^\eta \bar C^{\mu\nu}){\cal B}_{\nu\eta} + B \partial^\mu \bar C_2 + f^\mu B_2 + {\cal B}^{\mu\nu} \bar F_\nu
 \nonumber\\ &+& \bar F^\mu B_3 - (\partial^\mu \bar\beta^\nu - \partial^\nu \bar\beta^\mu) f_\nu \Bigr ].
\end{eqnarray}
As a consequence, the action integral ($S = \int d^6x {\cal L}_{(B, {\cal B})}$) remains invariant for the physically well-defined fields 
of the theory which vanish at infinity.

Corresponding to the symmetry transformations (10), we have a set of anti-co-BRST (i.e. anti-dual-BRST) symmetry transformations ($s_{ad}$):
\begin {eqnarray}
&&s_{ad} A_{\mu\nu\eta} = \frac{1}{2}\; \varepsilon_{\mu\nu\eta\kappa\rho\sigma} \partial^{\kappa} C^{\rho\sigma}, 
\quad s_{ad}  C_{\mu\nu} = - \bigl(\partial_\mu \beta_\nu - \partial_\nu \beta_\mu \bigr), \quad 
s_{ad} \beta_\mu = -\partial_\mu  C_2, \nonumber\\ &&s_{ad}  C_1 =  B, \quad s_{ad} {\bar C}_{\mu\nu} = {\bar {\cal B}}_{\mu\nu},
 \quad s_{ad} {\bar\beta}_\mu = \bar f_\mu, \quad s_{ad} {\bar C}_2 = -B_2, \quad s_{ad} {\bar C}_1 = B_3, \nonumber\\
&&s_{ad} \phi_\mu^{(2)} = F_\mu, \quad s_{ad} \bar F_\mu = - \partial_\mu B_3, \quad s_{ad} f_\mu = - \partial_\mu B,
 \quad s_{ad}  {\cal B}_{\mu\nu} = \partial_\mu F_\nu -  \partial_\nu F_\mu, \nonumber\\
&&s_{ad}\; [\partial^{\eta} A_{\eta\mu\nu}, \;\phi_\mu^{(1)},\; {\bar B}_{\mu\nu},\; {\bar {\cal B}}_{\mu\nu},\;
B, \;B_1,\; B_2,\; B_3,\; C_2,\; \bar f_\mu,\; F_\mu] \;= \;0.
\end{eqnarray}
The decisive features of the (anti-)dual-BRST [i.e. (anti-)co-BRST] symmetry transformations are as listed below:
\begin{itemize}
\item  Both are off-shell nilpotent of order two (i.e. $s^2_{(a)d} = 0$).
\item Both symmetries leave the total gauge-fixing terms:
\begin{eqnarray}
B^{\mu\nu}\Bigl(\partial^\eta A_{\eta\mu\nu} + \frac{1}{2} \Big[\partial_\mu \phi_\nu^{(1)} 
- \partial_\nu \phi_\mu^{(1)}\Big]\Bigr) - \frac{1}{2}\;B^{\mu\nu}B_{\mu\nu},\nonumber\\
- {\bar B}^{\mu\nu}\Bigl(\partial^\eta A_{\eta\mu\nu} - \frac{1}{2} \Big[\partial_\mu \phi_\nu^{(1)} 
- \partial_\nu \phi_\mu^{(1)}\Big]\Bigr) - \frac{1}{2}\; {\bar B}^{\mu\nu} {\bar B}_{\mu\nu},
\end{eqnarray}
invariant as $s_{(a)d} \;(\partial^{\eta} A_{\eta\mu\nu}) = 0, \; s_{(a)d}\;(\phi_\mu^{(1)}) = 0, \; s_{(a)d}\;(B_{\mu\nu}) = 0$ and
$s_{(a)d} \;({\bar B}_{\mu\nu}) = 0$.
\item Both symmetries are absolutely anticommuting (i.e. $s_d s_{ad} + s_{ad} s_d = 0$) on the constrained hypersurface 
defined by the following field equations:
\begin{eqnarray}
{\cal B}_{\mu\nu} + \bar {\cal B}_{\mu\nu} = \partial_\mu \phi_\nu^{(2)} - \partial_\nu \phi_\mu^{(2)}, \qquad
f_\mu + F_\mu = \partial_\mu C_1, \qquad \bar f_\mu + \bar F_\mu = \partial_\mu \bar C_1, 
\end{eqnarray} 
on the 6D Minkowskian flat spacetime manifold. In the above, the first entry has emerged from (8) and (9) due to the Euler-Lagrange equations of motion. 
\item The gauge-fixing term $(\partial^\eta A_{\eta\mu\nu})$, for the Abelian 3-form field $A_{\mu\nu\eta}$, 
owes its origin to the co-exterior 
derivative $\delta = - * d *$ because $\delta A^{(3)} = (1/2!)(dx^\mu \wedge dx^\nu) (\partial^\eta A_{\eta\mu\nu})$ 
produces it  on the 6D 
spacetime manifold. This term remains invariant under $s_{(a)d}$.
\item The Lagrangian density ${\cal L}_{(\bar B, \bar {\cal B})}$ transforms, under $s_{ad}$, as:
\begin {eqnarray} 
s_{ad} {\cal L}_{(\bar B, \bar {\cal B})} &=& - \partial_\mu \Bigl [-(\partial^\mu C^{\nu\eta} + \partial^\nu C^{\eta\mu}
 + \partial^\eta C^{\mu\nu})\bar {\cal B}_{\nu\eta} + \bar {\cal B}^{\mu\nu} F_\nu + B_2 \partial^\mu C_2 + F^\mu B_3
 \nonumber\\ &+& \bar f^\mu B + (\partial^\mu \beta^\nu - \partial^\nu \beta^\mu) \bar f_\nu \Bigr ],
\end{eqnarray}
which is the analogue  of (11). As a consequence,  the action integral of the theory remains invariant under the transformations 
$s_{ad}$. Out of the (anti-)co-BRST transformations, at least, one is certainly the analogue of the co-exterior derivative.
\item Under the (anti-)BRST symmetry transformations, the total kinetic term of (8) and (9) remains invariant because $s_{(a)b} {\cal B}_{\mu\nu} =0,
s_{(a)b} \bar {\cal B}_{\mu\nu} = 0, s_{(a)b} H_{\mu\nu\eta\kappa} = 0, s_{(a)b} \phi_\mu^{(2)} = 0$. Thus, there is a distinct difference
between $s_{(a)b}$ and $s_{(a)d}$.
\end{itemize}

\noindent
\section {Bosonic symmetry}
It is obvious, from the preceding sections, that we have four nilpotent ($s_{(a)b}^2 = 0, s_{(a)d}^2 = 0$) symmetries in the theory.
We have also shown that the (anti-)BRST and (anti-)co-BRST symmetries anticommute (i.e. $s_b s_{ab} + s_{ab}s_b = 0,
\; s_d s_{ad} + s_{ad}s_d = 0$)
separately and independently.
Furthermore, the following anticommutation relations \footnote {The algebra (16) is true for all the  fields
except $\beta_\mu, \bar \beta_\mu, \phi^{(1)}_\mu$ and $\phi^{(2)}_\mu$. It can be verified that $\{s_b, s_{ad}\} \bar\beta_\mu = \partial_\mu (B_1 - B_3)$,
$\{s_{ab}, s_d\} \beta_\mu = \partial_\mu (B_1 + B_3)$, $\{s_b, s_{ad}\} \phi^{(1)}_\mu = - \partial_\mu B, 
\;\{s_b, s_{ad}\} \phi^{(2)}_\mu = -\partial_\mu B,$ 
$\{s_{ab}, s_d \} \phi^{(1)}_\mu = \partial_\mu B_2$ and  $\{s_{ab}, s_d \} \phi^{(2)}_\mu = - \partial_\mu B_2$.
Thus, it is clear that the transformations $s_b$ and $s_{ad}$ are anticommuting only upto
the $U(1)$ vector gauge transformations. Similarly, the nilpotent transformations $s_d$ and $s_{ab}$ are {\it also} anticommuting only upto a $U(1)$
vector gauge transformation. The absolute anticommutativity property (i.e. $\{s_b, s_{ad}\} = 0, \{s_d, s_{ab}\} = 0$) is 
true for the rest of the fields of the theory. In other words, $s_b$ and $s_{ad}$ (as well as $s_d$ and $s_{ab}$) are independent transformations
upto a U(1) gauge transformation.
Furthermore, the anticommutators $\{s_d, \; s_{ab}\}$ and $\{s_b, \; s_{ad}\}$
{\it do not} define the bosonic transformations like $s_\omega$ and $s_{\bar \omega}$.}
\begin{eqnarray}
\{s_b, s_{ad}\} = 0,\qquad  \quad \{s_{ab}, s_{d}\} = 0,
\end{eqnarray}   
are true upto a $U(1)$ vector gauge transformation. The remaining non-vanishing anticommutation relations define the bosonic
transformations ($s_\omega, s_{\bar \omega}$) in the operator form. These transformations are succinctly expressed as:
\begin{eqnarray}
\{s_b, s_d\} = s_\omega,\qquad \quad \{s_{ab}, s_{ad}\} = s_{{\bar \omega}}.
\end{eqnarray}   
It turns out that the above transformations are {\it symmetry} transformations of the theory.

Let us first take the transformations $s_\omega$. Under this continuous symmetry transformations $s_\omega$, the fields of the theory transform as:
\begin{eqnarray}
&&s_\omega A_{\mu\nu\eta} = \frac {1}{2}\; \varepsilon_{\mu\nu\eta\kappa\rho\sigma}\partial^\kappa B^{\rho\sigma} + \Bigl(\partial_\mu {\cal B}_{\nu\eta} 
+ \partial_\nu {\cal B}_{\eta\mu} + \partial_\eta {\cal B}_{\mu\nu}\Bigr), \quad s_\omega \beta_\mu = \partial_\mu B, \nonumber\\
&& s_\omega C_{\mu\nu} = - (\partial_\mu f_\nu - \partial_\nu f_\mu), \quad s_\omega \bar C_{\mu\nu} = \partial_\mu \bar F_\nu - \partial_\nu \bar F_\mu, 
\quad s_\omega \bar \beta_\mu  = \partial_\mu B_2, \nonumber\\
&&s_\omega [ B, B_1, B_2, B_3, C_1, \bar C_1, C_2, \bar C_2, \phi_\mu^{(1)}, \phi_\mu^{(2)}, f_\mu, 
\bar f_\mu, F_\mu, \bar F_\mu, B_{\mu\nu}, \bar B_{\mu\nu},{\cal B}_{\mu\nu}, \bar {\cal B}_{\mu\nu}] = 0. 
\end{eqnarray} 
In the above, we have taken into account only the fields of the Lagrangian density (8). The latter itself transforms, under $s_\omega$, as follows:  
\begin{eqnarray}
s_\omega {\cal L}_{(B, {\cal B})} &=& \partial_\mu \Bigl[(\partial^\mu {\cal B}^{\nu\eta} + \partial^\nu {\cal B}^{\eta\mu}
+ \partial^\eta {\cal B}^{\mu\nu})B_{\nu\eta} - (\partial^\mu B^{\nu\eta} + \partial^\nu B^{\eta\mu}
+ \partial^\eta B^{\mu\nu}){\cal B}_{\nu\eta} \nonumber\\
&+& B_2 \partial^\mu B - B \partial^\mu B_2 + (\partial^\mu f^\nu - \partial^\nu f^\mu) \bar F_\nu 
+  (\partial^\mu \bar F^\nu - \partial^\nu \bar F^\mu) f_\nu   \Bigr],
\end{eqnarray}   
which establishes the fact that the action integral $(\int d^6x {\cal L}_{(B, {\cal B})})$ remains 
invariant under the  
transformations $s_\omega$. As a consequence, $s_\omega$ is truly a symmetry transformation.

The transformations for the fields of the Lagrangian density  ${\cal L}_{(\bar B, \bar {\cal B})}$ [cf.(9)] are as follows under the 
bosonic symmetry transformation $(s_{\bar \omega})$:
\begin{eqnarray}
&&s_{\bar \omega} A_{\mu\nu\eta} = \frac {1}{2}\; \varepsilon_{\mu\nu\eta\kappa\rho\sigma}\partial^\kappa \bar B^{\rho\sigma} 
+ \Bigl(\partial_\mu \bar {\cal B}_{\nu\eta} 
+ \partial_\nu \bar {\cal B}_{\eta\mu} + \partial_\eta \bar {\cal B}_{\mu\nu}\Bigr), \qquad s_{\bar\omega} \beta_\mu = - \partial_\mu B, \nonumber\\
&& s_{\bar \omega} C_{\mu\nu} = - (\partial_\mu F_\nu - \partial_\nu F_\mu), 
\qquad s_{\bar \omega} \bar C_{\mu\nu} = \partial_\mu \bar f_\nu - \partial_\nu \bar f_\mu, 
\qquad s_{\bar \omega} \bar \beta_\mu  = - \partial_\mu B_2, \nonumber\\
&&s_{\bar \omega} [ B, B_1, B_2, B_3, C_1, \bar C_1, C_2, \bar C_2, \phi_\mu^{(1)}, \phi_\mu^{(2)}, f_\mu, 
\bar f_\mu, F_\mu, \bar F_\mu, B_{\mu\nu}, {\bar B}_{\mu\nu}, {\cal B}_{\mu\nu}, {\bar {\cal B}}_{\mu\nu}] = 0. 
\end{eqnarray} 
It can be verified that the Lagrangian density ${\cal L}_{(\bar B, \bar {\cal B})}$ transforms, under the above transformations 
$s_{\bar \omega}$, to a total spacetime derivative, as given below:
\begin{eqnarray}
s_{\bar \omega} {\cal L}_{(\bar B, \bar {\cal B})} &=& - \partial_\mu\Bigl[(\partial^\mu \bar {\cal B}^{\nu\eta} + \partial^\nu \bar {\cal B}^{\eta\mu}
+ \partial^\eta \bar {\cal B}^{\mu\nu})\bar B_{\nu\eta} - (\partial^\mu \bar B^{\nu\eta} + \partial^\nu \bar B^{\eta\mu}
+ \partial^\eta \bar B^{\mu\nu}) \bar{\cal B}_{\nu\eta}\nonumber\\
&+& B_2 \partial^\mu B - B \partial^\mu B_2 + (\partial^\mu \bar f^\nu - \partial^\nu \bar f^\mu)F_\nu 
+ (\partial^\mu  F^\nu - \partial^\nu F^\mu) \bar f_\nu \Bigr],
\end{eqnarray}  
which shows that the action integral $(\int d^6x {\cal L}_{(\bar B, \bar {\cal B})})$ remains invariant under $s_{\bar \omega}$.

We wrap this section with the remark that, even though, the  bosonic symmetry transformations $s_\omega$ and $s_{\bar \omega}$ 
look different in their appearance,
actually, they are connected with each-other. In fact, it is straightforward to note that, if we exploit the CF-type restrictions
of (7) and (14), it can be readily verified that 
\begin{eqnarray}    
s_\omega + s_{\bar \omega} = 0 \qquad \Longrightarrow  \qquad (s_\omega + s_{\bar \omega})\;\Phi = 0,
\end{eqnarray}
where the generic field $\Phi = A_{\mu\nu\eta}, C_{\mu\nu}, \bar C_{\mu\nu}, \beta_\mu, \bar \beta_\mu$ corresponds to all the fields that
transform non-trivially under (18) and (20). We conclude that, out of the two bosonic symmetry transformations  $s_\omega$ and $s_{\bar \omega}$, only one 
bosonic symmetry transformation is independent.

\noindent 
\section {Ghost and discrete  symmetries}
The Lagrangian densities ${\cal L}_{(B, {\cal B})}$ and ${\cal L}_{(\bar B, \bar {\cal B})}$ have a part that is 
dependent on the (anti-) ghost fields and the other part contains fields that have ghost number equal to zero. The ghost part 
of the Lagrangian densities respects the following global scale transformations:
\begin{eqnarray}
&&C_{\mu\nu} \to e^{+\Omega} C_{\mu\nu}, \quad \bar C_{\mu\nu} \to e^{- \Omega} \bar C_{\mu\nu}, \quad \beta_\mu \to e^{+2\Omega}\beta_\mu,\quad
 \bar \beta_\mu \to e^{-2\Omega} \bar \beta_\mu,\quad C_2 \to e^{+3\Omega}C_2,\nonumber\\
&&\bar C_2 \to e^{-3\Omega}\bar C_2, \quad C_1 \to e^{+\Omega} C_1, \quad 
\bar C_1 \to e^{-\Omega} \bar C_1, \quad B \to e^{+2\Omega} B,\quad B_2 \to e^{-2\Omega}B_2,\nonumber\\
&& f_\mu \to e^{+\Omega} f_\mu, \quad  \bar f_\mu \to e^{-\Omega} \bar f_\mu,\quad  F_\mu \to e^{+\Omega} F_\mu, \quad  
\bar F_\mu \to e^{-\Omega} \bar F_\mu,
\end{eqnarray}  
where the numbers $(\pm1, \pm 2, \pm3)$ in the exponentials denote the ghost numbers of the corresponding ghost and anti-ghost fields and $\Omega$ is a 
spacetime independent global scale parameter. It should be pointed out that, under the ghost-scale transformations, the rest of the fields of   
${\cal L}_{(B, {\cal B})}$ and ${\cal L}_{(\bar B, \bar {\cal B})}$  transform as 
\begin{eqnarray}
&&A_{\mu\nu\eta} \to A_{\mu\nu\eta}, \quad B_{\mu\nu} \to B_{\mu\nu}, \quad {\cal B}_{\mu\nu} \to {\cal B}_{\mu\nu},\quad
B_1 \to B_1,\quad B_3 \to B_3,\nonumber\\
&&\bar B_{\mu\nu} \to \bar B_{\mu\nu}, \qquad \bar {\cal B}_{\mu\nu} \to \bar {\cal B}_{\mu\nu}, \qquad \phi^{(1)}_\mu \to \phi^{(1)}_\mu , 
\qquad \phi^{(2)}_\mu \to \phi^{(2)}_\mu.  
\end{eqnarray} 
This is due to the fact that the ghost number for the above fields is zero.
The infinitesimal version of the ghost scale transformations (23) (for $\Omega = 1$) is 
\begin{eqnarray}
&&s_g C_{\mu\nu} = + C_{\mu\nu}, \qquad s_g \bar C_{\mu\nu} = - \bar C_{\mu\nu}, \qquad s_g \beta_\mu = +2\beta_\mu,\qquad
s_g \bar \beta_\mu  = -2 \bar \beta_\mu,\nonumber\\
&&  s_g C_2 = +3C_2, \quad s_g \bar C_2 = -3 \bar C_2, \quad s_g C_1 = + C_1, \quad 
s_g \bar C_1 = - \bar C_1, \quad s_g B =+2 B,\nonumber\\
&& s_g B_2 = -2 B_2, \quad s_g f_\mu = + f_\mu, \quad  s_g \bar f_\mu = - \bar f_\mu,\quad  s_g F_\mu = + F_\mu, \quad  
s_g \bar F_\mu = - \bar F_\mu,
\end{eqnarray}  
and the infinitesimal version of (24) is
\begin{eqnarray}
s_g \Phi = 0, \qquad \qquad \Phi = A_{\mu\nu\eta},  B_{\mu\nu}, {\cal B}_{\mu\nu}, \bar B_{\mu\nu}, \bar {\cal B}_{\mu\nu},
\phi^{(1)}_\mu, \phi^{(2)}_\mu, B_1, B_3.  
\end{eqnarray}
These infinitesimal transformations (25) and (26) would play very important roles in the 
derivation of the algebraic structures that we shall discuss in the next section.

In addition to the above ghost-scale transformations, there exists the following discrete transformations for
the non-ghost sector of the Lagrangian densities (8) and (9): 
\begin{eqnarray}
A_{\mu\nu\eta} \to \pm  \frac{i}{3!}\; \varepsilon_{\mu\nu\eta\kappa\rho\sigma}\; A^{\kappa\rho\sigma}, \quad B_{\mu\nu} \to \pm i {\cal B}_{\mu\nu},
\quad {\cal B}_{\mu\nu} \to \pm i B_{\mu\nu}, \quad \bar B_{\mu\nu} \to \pm i \bar {\cal B}_{\mu\nu},\nonumber\\
\quad \bar {\cal B}_{\mu\nu} \to \pm i \bar B_{\mu\nu}, \quad \phi^{(1)}_\mu \to \pm i \phi^{(2)}_\mu, 
\quad \phi^{(2)}_\mu \to \pm i \phi^{(1)}_\mu, \quad B_1 \to \pm i B_3, \quad B_3 \to \pm i B_1, 
\end{eqnarray}
under which, the non-ghost part of the Lagrangian densities ${\cal L}_{(B, {\cal B})}$ and ${\cal L}_{(\bar B, \bar {\cal B})}$
remains invariant. Furthermore, under the discrete transformations given below: 
\begin{eqnarray}
&&C_{\mu\nu} \to \pm i \bar C_{\mu\nu}, \quad \bar C_{\mu\nu} \to \pm i C_{\mu\nu}, \quad \beta_\mu \to \pm i \bar \beta_\mu,\quad
 \bar \beta_\mu \to \mp i \beta_\mu,\quad C_2 \to \pm i \bar C_2,\nonumber\\
&&\bar C_2 \to \pm i C_2, \quad C_1 \to \mp i \bar C_1, \quad 
\bar C_1 \to \mp i C_1, \quad B \to \mp i B_2,\quad B_2 \to \pm i B,\nonumber\\
&& f_\mu \to \pm i \bar F_\mu, \quad  \bar F_\mu \to \pm i f_\mu,\quad  \bar f_\mu \to  \pm i F_\mu, \quad  
F_\mu \to \pm i \bar f_\mu,
\end{eqnarray} 
the ghost part of the Lagrangian densities ${\cal L}_{(B, {\cal B})}$ and ${\cal L}_{(\bar B, \bar {\cal B})}$ remains invariant.
The above discrete transformations (27) and (28) would play very important roles in the determination of the algebraic structures
amongst the symmetry transformations and its connection with the algebra of de Rham cohomological 
operators of differential geometry (cf. Sec. 6).

\noindent
\section {Algebraic structures}
It is clear, from the preceding sections, that there exist {\it six} continuous symmetries and {\it one} discrete symmetry in the theory.
The operator form of the continuous symmetry transformations obey the following algebraic structures, namely;
\begin{eqnarray}
&&s^2_{(a)b} = 0, \qquad s^2_{(a)d} = 0,\qquad  \{ s_b , s_{ab}\} = 0,\qquad \{ s_d , s_{ad}\} = 0, 
\qquad \{ s_b , s_{ad}\} = 0, \nonumber\\ 
&&  [s_g, s_b] = + s_b, \;\;\qquad
[s_g, s_{ab}] = - s_{ab}, \;\;\qquad [s_g, s_d] = - s_d, \;\;\qquad  [s_g, s_{ad}] = +s_{ad},\nonumber\\
&& \{ s_{ab} , s_d\} = 0, \quad \{ s_b , s_d\} = s_\omega = - \{ s_{ab} , s_{ad}\}, \quad
[ s_\omega , s_r ] = 0, \quad r = g, b, ab, d, ad. 
\end{eqnarray}
In the above, the infinitesimal version of transformations (3), (5), (10), (12), (18), (25) and (26) play crucial roles.
A close look at the above algebra shows that this algebra is the algebra of de Rham cohomological operators
($d,\delta,\Delta$) of differential geometry.

To corroborate the above statement, let us recall that  the set of de Rham cohomological operators of differential geometry
obey the following algebra [23-25]
\begin{eqnarray}
d^2 = 0, \quad \delta^2 = 0, \quad \{d, \delta\} = \Delta = (d + \delta)^2,\quad [\Delta, \delta] = 0, 
\quad [\Delta, d] = 0, \quad \delta = - *d*,
\end{eqnarray} 
where $*$ is the Hodge duality operator and $\Delta$ is the Casimir operator for the whole algebra.
The (co-)exterior derivatives (lower)raise the degree of a form by one on which they operate. In contrast,
the degree of a form remains intact when it is operated upon by $\Delta$.

An accurate comparison of (29) and (30) establishes the fact that the sets ($s_b, s_d, s_\omega$) as well as ($s_{ab}, s_{ad}, - s_\omega$)
are the analogue of the de Rham cohomological operators ($d, \delta, \Delta$). The central row of equation (29) proves the fact that the
ghost number of a field increases by one when it is operated upon by $s_b$ and $s_{ad}$. On the contrary, the  ghost number of a field 
decreases by one when $s_d$ and $s_{ab}$ act on it. Thus, the ghost number plays the role of the degree  
of a form when we compare the symmetry operators of our present  theory and the cohomological operators (of differential geometry), respectively.

Finally, we make statements on the analogue of the Hodge duality $*$ operator in the language of the symmetry transformations.
It turns out that the discrete symmetry transformations listed in (27) and (28), combined together, lead to the derivation of the following
relationship between the continuous symmetry transformation $s_{(a)b}$ and $s_{(a)d}$, namely; 
\begin{eqnarray}
s_{(a)d} \;= \;\pm \;*\;s_{(a)b}\;*, 
\end{eqnarray} 
where $*$ corresponds to the transformations listed in (27) and (28). The $(+)-$ signs, 
in the above, are dictated by the two successive operations of the discrete transformations (27) and (28) 
on a generic field $\Phi$ of the theory. This statement can be mathematically expressed in a succinct form 
as follows (see, e.g. [28] for details on duality):
\begin{eqnarray}
*\;(*\;\Phi) = \pm\;\Phi.
\end{eqnarray} 
One can explicitly check that  the $+$ sign, in the above, is true only for the four fields. These fields are $\Phi = \beta_\mu, \bar \beta_\mu, B, B_2$.
Rest of the fields correspond to ($-$) sign in (32).

\noindent
\section {Conclusions}
In our present investigation, we have shown the existence of local, off-shell nilpotent
and absolutely anticommuting (anti-)co-BRST symmetry transformations in the
context of a 6D Abelian 3-form gauge theory. One of the decisive features of the above symmetry transformations is the fact that
the total gauge-fixing term, owing its origin to the co-exterior derivative, remains invariant. Thus, one of
the (anti-)co-BRST symmetry transformations is an analogue of the co-exterior derivative.
On the contrary, it is the kinetic
term that remains invariant under the  nilpotent (anti-)BRST symmetry transformations. Thus, the off-shell nilpotent BRST transformation is an analogue
of the nilpotent exterior derivative because the BRST invariant kinetic term has its origin in this cohomological operator.

We have demonstrated that a suitable anticommutator of the above (anti-)BRST and (anti-)co-BRST symmetry transformations leads to the definition of
a {\it single} independent bosonic symmetry in the theory, under which, the (anti-)ghost fields transform to the $U(1)$ 
vector as well as tensor gauge transformations [cf. (18)].
We have also shown that the bosonic symmetry ($s_\omega$) commutes with all the rest of the symmetry transformations of the theory. As a consequence,
it is the Casimir operator for the whole algebra and 
it behaves like the Laplacian operator of differential geometry. Thus, in our present work, we have clearly provided the physical realizations
of the cohomological operators ($d, \delta, \Delta$).

One of the novel observations of our present endeavor is that the anticommutators $\{s_b, s_{ad}\}$ and $\{s_{ab}, s_d\}$ 
are {\it  not} absolutely zero for the specific fields of theory. In fact, as it turns out, the absolute
anticommutativity of the above operators is true only upto
the $U(1)$ gauge transformations. This observation is totally different from our earlier works on the free 4D Abelian 2-form [20,21] as well as 2D 
(non-)Abelian 1-form (free as well as interacting) gauge theories [18,19,17] where we have absolute anticommutativity between $s_d$ and $s_{ab}$
as well as $s_b$ and $s_{ad}$. We very strongly feel that such kind of novel observations will appear in the context of BRST
analysis of the higher dimensional ($D>4$) and higher $p$-form ($p \geq 3$) gauge theories that are important in the context of (super)string theories.

In our present endeavor, we have discussed only the symmetry transformations. We plan to calculate all the generators for the continuous 
symmetry transformations of the present theory and establish that the present theory is a model for the Hodge theory. 
Further, we also wish to study its topological features. We feel that the present theory might turn out to be a cute
field theoretic model for a topological field theory as we have been able to demonstrate in the case of the 
free 2D (non-)Abelian 1-form gauge theories [17]. 
These are some of the issues that are presently under intensive investigation and a longer version of our present 
work will be reported in our future publications. In this context, it is gratifying to 
state that we have already accomplished some of the above goals in  [29]. \\

\noindent
{\bf Acknowledgement:}
\noindent 
One of us (RK) would like to thank University Grant Commission (UGC), 
Government of India, New Delhi, for financial support.\\


\begin{thebibliography}{99}
\bibitem{gsw}   M. B. Green, J. H. Schwarz, E. Witten, {\it Superstring Theory} Vols 1 and 2\\
                (Cambridge University Press: Cambridge, 1987)
\bibitem{p}     J. Polchinski, {\it String Theory} Vols 1 and 2\\
                (Cambridge University Press: Cambridge, 1998)
\bibitem{lt}    D. Lust, S. Theisen, {\it Lectures in String Theory} (Springer-Verlag: New York, 1989)
\bibitem{abl}   T. J. Allen, M. J. Bowick, A. Lahiri, Mod. Phys. Lett. A {\bf 6}, 559 (1991)
\bibitem{al1}   A. Lahiri, Mod. Phys. Lett. A {\bf 8}, 2403  (1993)
\bibitem{trg}   T. R. Govindarajan, J. Phys. G {\bf 8}, 117 (1982)
\bibitem{ft1}   D. Z. Freedman, P. K. Townsend, Nucl. Phys. B {\bf 177}, 282 (1981)
\bibitem{hls}   E. Harikumar, A. Lahiri, M. Sivakumar, Phys. Rev. D {\bf 63}, 105020 (2001)
\bibitem{al2}   A. Lahiri, Phys. Rev. D {\bf 55}, 5045  (1997)
\bibitem{hen}   M. Henneaux, V. E. R. Lemes, C. A. G. Sasaki, S. P. Sorella, O. S. Ventura,\\
                L. C. Q. Vilar, Phys. Lett. B {\bf 410}, 195 (1997)
\bibitem{srm1}  S. Gupta, R. Kumar, R. P. Malik, Eur. Phys. J. C {\bf 70}, 491  (2010) 
\bibitem{ksm1}  S. Krishna, A. Shukla, R. P. Malik, Int. J. Mod. Phys. A {\bf 26},  4419 (2011)
\bibitem{km1}   R. Kumar, R. P. Malik, Euro. Phys. Lett. {\bf 94}, 11001 (2011)
\bibitem{km2}   R. Kumar, R. P. Malik, Eur. Phys. J. C {\bf 71}, 1710  (2011) 
\bibitem{cbiz}  C. Bizdadea, Phys. Rev. D {\bf 53}, 7138 (1996)
\bibitem{ksm2}  S. Krishna, A. Shukla, R. P. Malik, Mod. Phys. Lett. A {\bf 26}, 2739  (2011)
\bibitem{rpm1}  R. P. Malik, Int. J. Mod. Phys. A {\bf 22}, 3521 (2007)
\bibitem{rpm2}  R. P. Malik, Mod. Phys. Lett. A {\bf 19}, 2721 (2004)
\bibitem{rpm3}  R. P. Malik, Mod. Phys. Lett. A {\bf 16}, 477 (2001)      
\bibitem{gm1}   S. Gupta, R. P. Malik, Eur. Phys. J. C {\bf 58}, 517  (2008)
\bibitem{gkm}   S. Gupta, R. Kumar, R. P. Malik, Eur. Phys. J. C  {\bf 65}, 311  (2010)
\bibitem{bm1}   See, e.g., R. P. Malik, Int. J. Mod. Phys. A {\bf 19}, 5663 (2004) 
\bibitem{egh}   T. Eguchi, P. B. Gilkey, A. Hanson, Phys. Rep. {\bf 66}, 213 (1980)
\bibitem{smm}   S. Mukhi, N. Mukanda, {\it Introduction to Topology, Differential Geometry and\\
                Group Theory for Physicists} (Wiley Eastern Private Limited, New Delhi, 1990)  
\bibitem{vh}    J. W. van Holten, Phys. Rev. Lett. {\bf 64}, 2863 (1990)
\bibitem{bm2}   L. Bonora, R. P. Malik, J. Phys. A: Math. Theor. {\bf 43}, 375403 (2010)  
\bibitem{rpm4}  R. P. Malik, Eur. Phys. J. C {\bf 60}, 457  (2009)  
\bibitem{dght}  S. Deser, A. Gomberoff, M. Henneaux, C. Teitelboim,  Phys. Lett. B {\bf 400}, 80 (1997) 
\bibitem{kksm}  R. Kumar, S. Krishna, A. Shukla, R. P. Malik, arXiv:1203.5519[hep-th]
\end{thebibliography}
\end{document}